\documentclass[a4paper,11pt]{article}
\usepackage{rotating}
\usepackage{amssymb} 
\usepackage{amsmath} 
\usepackage{graphicx}
\usepackage[flushmargin]{footmisc}

\author{A.~Noutsos\thanks{University of Manchester, Jodrell Bank Centre for Astrophysics, Alan Turing Building, Manchester M13 9PL, UK}, A.~Karastergiou\thanks{Astrophysics, University of Oxford, Denys Wilkinson Building, Keble Road,Oxford OX1 3RH, UK}
, M.~Kramer$^*$, S.~Johnston\thanks{Australia Telescope National Facility, CSIRO, P.O. Box 76, Epping, NSW 1710, Australia.}
 and B.~W.~Stappers$^*$}

\date{\today} 
\title{Phase-resolved Faraday rotation in pulsars}

\begin{document}

\maketitle

\begin{abstract} 
We have detected significant Rotation Measure variations for 9 bright pulsars, as a function of pulse longitude. An additional sample of 10 pulsars showed a rather constant RM with phase, yet a small degree of RM fluctuation is visible in at least 3 of those cases. In all cases, we have found that the rotation of the polarization position angle across our 1.4 GHz observing band is consistent with the $\lambda^2$ law of interstellar Faraday Rotation.  
We provide for the first time convincing evidence that RM variations across the pulse are largely due to interstellar scattering, although we cannot exclude that magnetospheric Faraday Rotation may still have a minor contribution; alternative explanations of this phenomenon, like erroneous de-dispersion and the presence of non-orthogonal polarization modes, are excluded. If the observed, phase-resolved RM variations are common amongst pulsars, then many of the previously measured pulsar RMs may be in error by as much as a few tens of rad m$^{-2}$. 
\end{abstract}

\section{Introduction}
Radio pulsars are generally highly polarised sources, and the
propagation of their radiation through the magneto-ionised
interstellar medium (ISM) is affected by Faraday rotation. Measurement of
the rotation measure (RM), combined with the dispersion measure (DM)
for each pulsar, provides a good estimate of the average magnetic-field strength in the direction of the line of sight to many of the
known pulsars. With the high sensitivity and broad bandwidths of
current observing instruments, it is possible to accurately measure
RMs within a single band, thus avoiding calibration issues of
observing with different receivers in different bands. Furthermore,
observing pulsars with high temporal resolution allows RM measurements
at different pulse longitudes, provided the signal to noise ratio of
the linearly polarised component is sufficient.

Ramachandran et al. (2004) presented a phase-resolved RM analysis for
PSR B2016+28. In an attempt to explain the variation in RM by
$\pm15$~rad~m$^{-2}$ across the pulse profile, they conducted a single-pulse analysis of the polarization. They found that there are clearly
two, almost orthogonal polarization modes across the pulse profile,
with slightly different spectra (as also postulated in Karastergiou et
al. 2005 and demonstrated in Smits et al. 2006). Superposition of
these modes results in a net change in the polarization position angle
with frequency, leading to the measurement of a variable RM across the
pulse.  Although the effect measured in their study rotates the position
angle (PA) with frequency, they concluded that it is not Faraday rotation
intrinsic to the pulsar magnetosphere. The authors clearly demonstrate that for PSR B2016+28 the position angle does not have a quadratic dependence on wavelength: i.e.~$\Delta{\rm PA}={\rm RM}\cdot\lambda^2$.

A very recent work by \cite{kar09} showed that even small amounts
of interstellar scattering can significantly deform the steep features of PA profiles (e.g.~orthogonal mode jumps), and can provide a simple explanation for the RM variations observed by Ramachandran et al. The latter is caused by the frequency dependence of the scattering constant, which causes a frequency-dependent amount of polarised intensity to be transfered from a given phase bin to later phases. Regions where the PA profile has a steep gradient are worst affected, since it is at those phases where the scattered portion of the polarised intensity has the most impact on the shape of the PA profile: flat profiles can only stay flat, but the kinks and wiggles in PA profiles will be distorted by the redistribution of the polarised intensity.  
Hence, Karastergiou noted that the polarization of core components, where PA profiles are steepest, is affected the most. The author concluded that scattering can smear orthogonal polarization modes so that they appear slightly less ``orthogonal''.   

In this paper we concentrate on pulsars for which we have recently
published RM values (e.g. Noutsos et al. 2008). Our aim is not only to
fit for pulse phase--resolved RMs, as was done in Ramachandran et al., but to
closely investigate whether the polarization position angles obey
$\lambda^2$ fits. We investigate the possible causes for this
particular PA frequency dependence and discuss implications on RM
measurements performed using different techniques.

\section{Observations and data analysis} 
The data used in our analysis were collected with the Parkes 64-m
radio telescope during two observing periods: in 2004, August 31 to
Septemeber 1, and in 2006, August 24 to 27. We observed at the
frequency band of the 20cm H-OH receiver, installed at the focus cabin
of the Parkes telescope. The receiver is equipped with a pair of
orthogonal, linear feeds that are sensitive to polarised emission with
an equivalent system flux density of 43 Jy at 20cm (\nocite{joh02}Johnston 2002). The 2004 data were
observed with a central frequency of 1.375 GHz and 256 MHz bandwidth,
while in 2006 we observed with a central frequency of 1.369 GHz and
256 MHz bandwidth.

The wide-band correlator (WBC) back end that was used in both
observations is capable of splitting the available bandwidth
into 1,024 frequency channels, on each of four polarizations, $I$, $Q$,
$U$ and $V$, and producing 10-s de-dispersed pulse subintegrations
with up to 2,048 phase bins across the pulse profile.

In Noutsos et al. (2008), we outline the method used to derive a
single RM value for each pulsar we observed. Here, we performed the
same analysis of the polarised emission in each longitude bin, across
the total pulse width. Our aim was to produce RM profiles as a
function of longitude for each observed pulsar. Prior to our analysis,
the raw data from the WBC were subjected to the
necessary process of RFI mitigation and polarization calibration. We
processed the calibrated data using the PSRCHIVE software package
\cite{hvm04}. 

It is expected that the plane of linearly polarised emission will be
rotated by the magnetised ISM it traverses on its way to the Earth:
this is the well-known interstellar Faraday effect. The amount of
rotation, $\Delta {\rm PA}$, for a given pulsar should be constant for
a specific frequency but scales as $1/f^2$ across the observing
band. By definition, the rotation between frequency $f$ and infinite
frequency gives the PA at frequency $f$:

\begin{equation} 
\label{eq:paeq2} 
{\rm PA}=c^2{\rm RM}\frac{1}{f^2} 
\end{equation} 

Since the RM is a physical property of the ISM, it is expected to be
constant for a given ISM configuration between the pulsar and the
Earth. Its value depends on the magnitude and direction of the
magnetic field, $\boldsymbol{B}$, along the propagation path of
the polarised emission, $C$; the free-electron density, $n_e$, along $C$,
and the length of $C$; it is given by the path integral

\begin{equation}
\label{eq:rmeq1} 
{\rm RM} \propto \int_{C}n_e\boldsymbol{B} \cdot d\boldsymbol{l} 
\end{equation} 
where $d\boldsymbol{l}$ is directed from the pulsar to the Earth.    

Apart from the pulsar distance, none of the remaining physical
quantities, on which RM depends, are directly measurable. However,
given a finite observing bandwidth, one can derive RM by fitting the
PA across the band with the quadratic function of equation~\ref{eq:paeq2}. The
PA of each bin $i$ in each frequency channel $j$ was calculated as
${\rm PA}_{i,j}=0.5\arctan{(U_{i,j}/Q_{i,j})}$. 

Having previously
developed the software capable of such quadratic fits (see
\nocite{njk+08} Noutsos et al. 2008 for details), we performed
bin-wise fitting across the longitude range inside the pulse and
derived a value of RM for each individual longitude bin. In order to
increase the s/n per frequency channel, $j$, before fitting across the
whole available bandwidth we averaged the frequency channels by a
factor 16. For a single averaging factor, as is the case in this
analysis, only the statistical error on each RM value can be
calculated, and the fitting routine gives the
best integer RM for each bin. Finally, we plotted pulse profiles of the total
flux density, the linear- and circular-polarization components, and
the PA and RM profiles across the pulse; we only plotted significant PA
values with ${\rm s/n} > 3.5$ and RM values with $\sigma_{\rm RM}<5$
rad m$^{-2}$.

\section{Results}
 \label{subsec:20cmdata}
\begin{figure}\begin{center} 
\vspace*{10pt}
\includegraphics[width=0.8\textwidth]{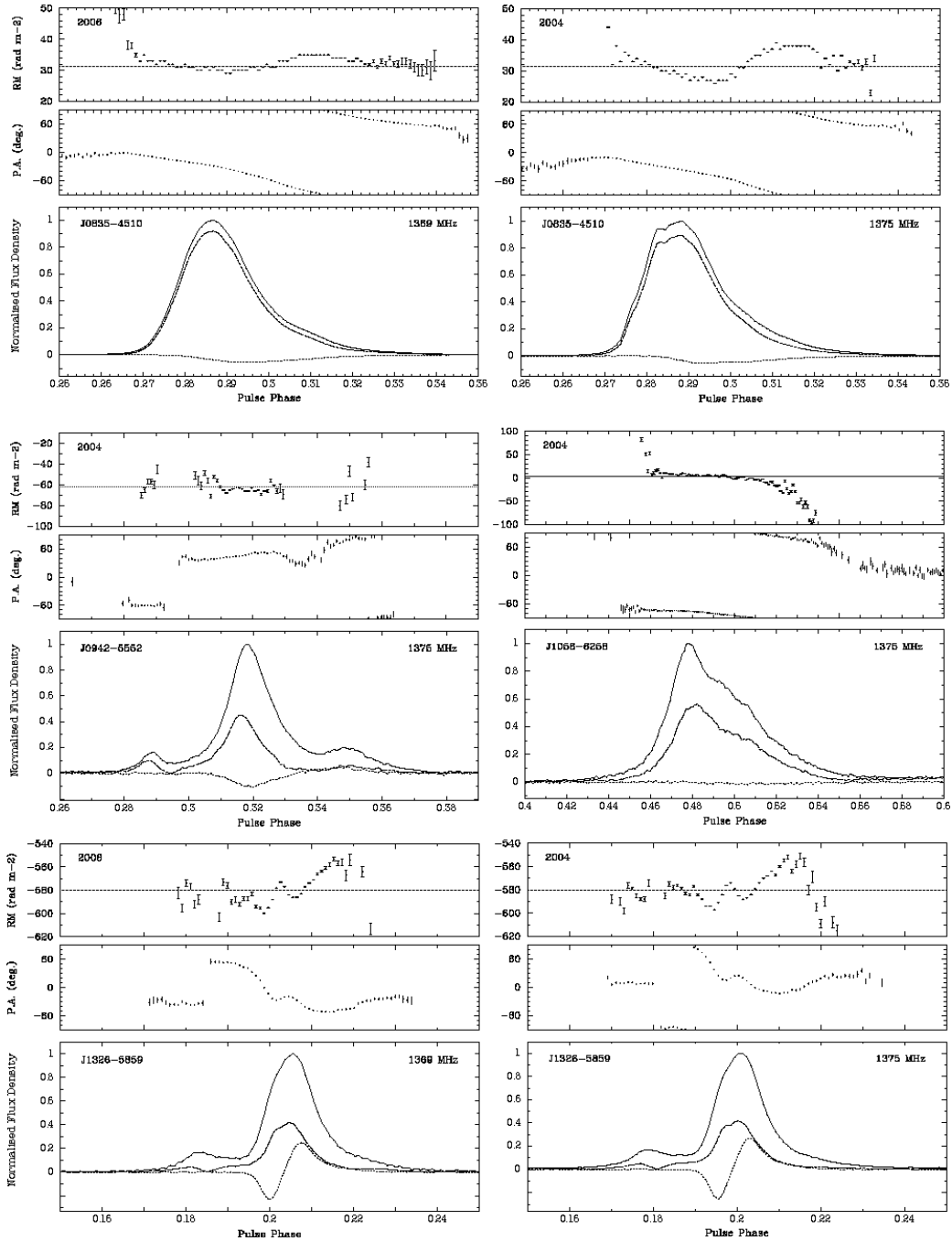}
\caption{\label{fig:largeRMvar1} Profiles of the RM as a function of pulse longitude (top panels), for 9 pulsars, observed at 20 cm with the Parkes telescope, that exhibit large RM variations across their pulse profile. The respective year of observation is shown at the top-left corner of each pulsar's RM profile. For two pulsars, namely the Vela pulsar and PSR J1326$-$5859, we had data from two observing sessions, in 2004 and 2006. The RM for each longitude bin was calculated from a quadratic fit to the PAs of that bin across the frequency band. Only RM values with $\sigma_{\rm RM}<5$ rad m$^{-2}$ are shown. The horizontal, gray line across the RM profiles shows the published RM value for each pulsar.
The middle panels show the phase-resolved PA profiles, calculated from the integrated Stokes profiles, $Q$ and $U$, across the respective frequency band. For each longitude bin, the PA was calculated as ${\rm PA}=0.5\arctan(U/Q)$. Only PAs corresponding to linear polarisation with ${\rm s/n}>3.5$ are shown. The bottom panels show the profile of the total flux density (solid lines) and the linear (dashed lines) and circular (dotted lines) polarization flux for each pulsar.}
\end{center}\end{figure}

The resulting 19 RM profiles are shown in Fig.~\ref{fig:largeRMvar1}
and Fig.~\ref{fig:smallRMvar1}. 9 pulsars show large RM variation
across the pulse and 10 pulsars show a rather constant RM profile. In
these plots, the bottom panel shows the polarization profile, where
solid lines denote total power; dashed lines, linear polarization; and
dotted lines, circular polarization. The middle panels show the
integrated PA profile and the top panels, the RM profile. In the interest of
characterisation, we have included a number of properties of the pulsar emission and the ISM in the direction of the 19 pulsars, in Table~\ref{tab:pulprop}.

\subsection{Pulsars with large RM variation}

{\bf PSR J0835$-$4510 (2004 \& 2006 data).} The Vela pulsar has a
highly polarised, yet relatively simple pulse profile in both the 2004
and 2006 data. Both profiles appear scattered.  The PA shows the
characteristic swing which has been used as a strong argument for the
rotating vector model (\nocite{rc69a}Radhakrishnan \& Cooke 1969). The most striking profile of Vela is its RM
profile, which shows a clear swing across the pulse that resembles a
sideways `S' (peak-to-peak variation $\approx 6$ rad m$^{-2}$). The
shape of the RM profile persists through both 2004 and 2006 data sets,
albeit the earlier of the two profiles displays a somewhat higher peak-to-peak variation of the RM ($\approx 13$ rad m$^{-2}$).

{\bf PSR J0942$-$5552 (2004 data).} The pulse profile of this pulsar
has 3 distinct components, with the peaks of linear polarization being
slightly offset from the corresponding ones in the total power. The PA
profile shows an orthogonal jump at the bridge emission between the
leading and central components and at the exact phase where $L$
vanishes. The RM profile shows a steep decline of high-s/n points
coincident with the leading end of the central component and an
equally steep increase, coincident with the leading component; this
behaviour occurs around the orthogonal PA jump. In both cases, the RM
changes by at least 20 rad m$^{-2}$. At the central peak's maximum the
RM fluctuates within a few rad m$^{-2}$.

{\bf PSR J1056$-$6258 (2004 data).}  The total-power profile of this
pulsar appears very slightly stretched toward later pulse phases. It
has nearly zero circular polarization but $\approx 50$\% linear. The
PA profile sweeps smoothly across the pulse, without showing any
breaks. The RM profile is roughly consistent with the published RM
around the the flux maximum, but it drops suddenly near the pulse edges,
giving it an overall shape of a swing across the entire pulse. \cite{kj06} performed a comparison between the 20cm and 10cm profiles of this pulsar and concluded that the linear polarization, and hence the PA profile, must undergo significant frequency evolution. They supported their argument by deriving a pulse-averaged RM value for PSR B1056$-$6258 that was much closer to the published one (6 rad m$^{-2}$), whereas if no evolution was assumed that value was $-1$ rad m$^{-2}$. However, this pulsar shows a peak-to-peak RM variation across the pulse of almost 100 rad m$^{-2}$ --- the highest in our sample --- which dwarfs any possible RM error due to profile evolution.

{\bf PSR J1243$-$6423 (2004 data).} The central pulse of this pulsar's
profile consists of two Gaussian components, whereas a third, much
weaker component is present at the far leading edge of the
profile. The circular polarization profile resembles an unsusual
double swing, with its minimum value being roughly coincident with the
maximum $L$. There is an orthogonal PA jump after the leading
component. The PA profile corresponding to the central pulse
components displays a full `S'-shaped swing, characterised by wiggles
often seen in PA profiles. The RM profile of this pulsar is extremely
complex: it shows a clear structure that departs from flatness, with a
leading swing followed by a sharp peak and a steep decline of the
RM. The peak-to-peak variation of the profile is $\approx 60$ rad
m$^{-2}$. The maximum RM ($\approx 200$ rad
m$^{-2}$) appears coincident with the inflexion point of the PA swing.

{\bf PSR J1326$-$5859 (2004 \& 2006 data).} This pulsar's profile is the sum of at least two components: a weak leading one and
a strong central one that leads into a trailing tail, perhaps
indicative of a third, very weak component. The linear polarization of
the central component is moderatly high, with more than 30\% of its
total flux being polarised, whereas the leading component is just over
20\% polarised. Even more interesting is the evident swing of the
circular polarization under the central component. Like in the case of
PSR J1157$-$6224, this may well be hinting at emission from near the
magnetic axis. In both epochs, the PA profile is broken by an
orthogonal jump near the peak of the leading component, at the exact
point where $L$ vanishes. At later phases, the PAs form an unusual
swing interrupted mid-way by a bulge that is coincident with the
maximum of the linear polarization. The RM profile clearly shows a
smooth functional dependence on phase that deviates from flatness;
like in the case of Vela, this dependence is exactly reproduced in both
2004 and 2006 data, which argues against this being an instrumental
effect. The form of the RM function resembles a double swing with 2
local minima separated by a local maximum. The trailing of the two
minima is exactly coincident with the peak in total power.

{\bf PSR J1453$-$6413 (2004 data).} The pulse profile of this
pulsar is a composition of a relatively weak leading component and a
much stronger central one. However, in the linear-polarization profile
these two components are of comparable intensity. The PA profile
swings along the entire central component but is disrupted by an
orthogonal jump that coincides with the merging point of the two
components' linear polarization, where $L=0$. The RMs across the pulse
show a moderate amount of scattering that prevents us from being
certain about the existence of a possible structure. On one hand,
there is a clear break in the RMs due to the jump in the PAs. On the
other hand, even away from the jump there seems to be a dip in the RMs
that is coincident with the peak of the central component's linear
polarization.

{\bf PSR J1644$-$4559 (2004 data).} The total power of this pulsar can
be described by a single Gaussian with an extended, trailing scattering
tail (see also analysis by \nocite{joh04}Johnston 2004). A hint at a second, very weak component can be seen near the
leading edge of the main pulse. The maximum of the linear polarization
occurs at slightly earlier phase than that of the total-power
maximum. In addition, this pulsar exhibits a swing in the circularly
polarised profile. The PA profile shows a more complex picture, in
comparison: it features two smooth bumps across the main pulse, which
are only interrupted by an approximately orthogonal jump, coincident
with the bridge emission between the main pulse and the weak leading
component; it also shows the obvious signs of scattering on
polarization shown in Li \& Han (2001)\nocite{lh03}. The RM profile
features a clear structure that comprises at least three local maxima
and an equal number of local minima; none of these extremes is exactly
coincident with either the total flux or the linearly polarised flux
extremes. The fluctuation of the RM across the pulse appears mildly
correlated to the PA profile.

{\bf PSR J1745$-$3040 (2004 data).} There are three, distinct, highly
polarised components in this pulsar's profile. \cite{jkm+08} performed multifrequency observations of this pulsar and noted the significant frequency evolution of its profile as well as the apparent increase of the linear polarization fraction with frequency. The PA profile
resembles an upside-down $\upsilon$ across the trailing and central
components, which is interrupted by an orthogonal jump occuring
between the trailing components and the leading one. The RM profile
shows a seemingly constant downwards slope for the most part of the
two trailing components but kinks downwards at the leading edge of the
central component. Above the leading component, the RMs are scattered,
although an upwards slope can perhaps be inferred from the 6 data points
coincident with this component. 

{\bf PSR J2048$-$1616 (2006 data).} Three Gaussian components can be
discerned in the total profile of this pulsar. The linear polarization
is strong in the leading and trailing of the three components but
relatively weak in the central one. The PA profile resembles an
`S'-shaped swing that rotates the PAs smoothly across the profile (see also \nocite{jkk+07}Johnston et al. 2007; \nocite{jkm+08}Johnston et al. 2008). The
RM is rather constant across the leading component but dips by
$\approx 8$ rad m$^{-2}$ at the central, least-polarised component. A
sudden rise of the RMs is observed on the trailing edge of the
trailing component, although the s/n of the linear polarization is
already low in that phase region.

\subsection{Pulsars with low RM variation}

{\bf PSR J0134$-$2937 (2006 data).} The PA profile of this pulsar is
flat everywhere, apart from the phase region around the pulse maximum,
where the profile is broken by an orthogonal jump. The jump is also
coincident with the narrowest of the two distinct components present in
the linearly polarised flux, the other being broad and coincident with
the long leading tail of the total flux. The RM profile of this pulsar
shares the same flatness, which seems to be disturbed by the presence
of the orthogonal jump. However, the RM fluctuation occurs only over a
few phase bins, and its magnitude is of the same order as at least
three other instances away from the orthogonal jump.

{\bf PSR J0738$-$4042 (2004 \& 2006 data).}  The polarization of this
pulsar was shown at two frequencies (1.4 and 3~GHz) in Karastergiou \&
Johnston (2006). The position angle shows two orthogonal-mode
jumps. The RM profile, although generally flat, is below the average
RM value in the leading part of the profile; it then turns above the
average RM value for the rest of the profile, after the first
orthogonal jump. In the dual frequency data of Karastergiou \&
Johnston, the frequency evolution of the position-angle profile is in
agreement with what is seen here: the first segment of the PA profile
has a different frequency dependence to the rest of the profile.

{\bf PSR J0742$-$2822 (2004 \& 2006 data).} This pulsar comprises a
complex, 7-component profile (see e.g.~\nocite{kra94}Kramer
1994; \nocite{kjm05}Karastergiou et al. 2005). Despite the complexity of the pulse itself, the PA and RM
profiles display a relatively simple dependence on phase, with the
former being a simple sweep across the profile and the latter being
rather flat. It is perhaps noteworthy that a comparison
between the profiles of 2004 and 2006 shows that the average RM in the earlier observations appears slightly higher than the published value, whereas the opposite is observed in 2006. The $\langle {\rm RM}\rangle$ difference between the two data sets is $\approx 5$ rad m$^{-2}$.

{\bf PSR J0907$-$5157 (2006 data).} The total-power and linear-polarization profiles of this pulsar have two components. The PA profile
is a smooth, continuous sweep, without any breaks, and one can notice
that it slopes slightly upwards, toward the trailing edge of the
pulse profile. The RM profile is flat and consistent within the errors
with the published RM.

{\bf PSR J1057$-$5226 (2006 data).} This young pulsar ($5.4\times 10^5$ y) is an interesting and a peculiar one, and one of a few
that are visible at all wavelengths (see e.g.~\nocite{vl72}Vaughan \& Large 1972; \nocite{ch83}Cheng \& Helfand 1983;
\nocite{mcb97}Migniani et al.~1997; \nocite{fbc+94}Fichtel et al.~1994). Even more interesting is its strong radio interpulse, which
is separated from the main pulse by roughly 150$^\circ$ and has the same energy as the main pulse over a wide range of radio
frequencies (\nocite{mhak76}McCulloch et al.~1976; Weltevrede \& Wright 2008 {\em MNRAS submitted}); it has been suggested that we observe emission from both magnetic poles
(\nocite{lm88}Lyne \& Manchester 1988; \nocite{wqxl06}Wang et al.~2006). Furthermore, the steepness of the PA swing for the main
pulse and the interpulse differs significantly, which has given rise to the suggestion that the corresponding emission is generated
at different heights in the magnetosphere (\nocite{wqxl06}Wang et al.~2006). Due to the higher degree of linear polarization ($\sim 90\%$) and the modest s/n of the pulse in the 2006 data, we chose to study only the main pulse for RM variations. The PA profile of the main pulse is flat, with no breaks, and shows rotation of no more than 60$^{\circ}$ across the entire pulse. The RM profile is also consistent with a constant RM. 

{\bf PSR J1157$-$6224 (2004 data).} This pulsar shows a simple total-power profile, with substantial amounts of linear ($>25$\%) and circular
polarization. 
The PA profile exhibits a smooth but
complex swing, broken only by an orthogonal jump half-way along the
trailing edge of the pulse.  The RM profile also swings, although the
edges are too noisy and the magnitude of the effect too small to draw a
certain conclusion.

{\bf PSR J1253$-$5820 (2006 data).} This pulsar's pulse profile can be
fitted with a single Gaussian component, and the same pulse shape is
shared by the linearly polarised flux, which accounts for a large
fraction of the total observable emission. The PA shows a smooth
variation throughout. The RM profile shows little deviation from
flatness, mainly at the leading edge.

{\bf PSR J1359$-$6038 (2004 data).} This pulsar's profile is $>70$\%
linearly polarised and $\approx 15$\% circularly polarised. The PA profile
is rather flat across the pulse. The RM profile is consistent with a
constant value for the most part of the profile, diverging slightly at
the tail end, similarly to PSR J1056$-$6258. However, at that point the s/n is greatly reduced so solid conclusions cannot be extracted.

{\bf PSR J1705$-$1906 (2006 data).} This is another highly polarised
pulsar with a main pulse composed of multiple components: despite having been classified as a three-component pulse, at least five components can be discerned in both the total power and the linear polarization
profiles in our data (\nocite{ran90}Rankin 1990; \nocite{gl98}Gould \& Lyne 1998). In addition, a much simpler, single-peaked interpulse is also present in this pulsar's profile, which is separated by roughly $180^\circ$ from the main pulse. At 20cm, the interpulse is almost 100\% linearly polarised, which has also been noted in previous work (see e.g.~\nocite{kj04}Karastergiou \& Johnston 2004). In this work, we chose to examine only the main pulse for RM variations, due to its larger width and higher s/n. The PA profile is a gradual sweep across the entire main pulse. The RM profile appears flat, on average, although a coherent jitter over a-few-bin scale is evident. The only possible departure from flatness appears to coincide with the leading outrider.

{\bf PSR J1807$-$0847 (2004 data).} Another case of a multi-component,
total-power profile, with low linear polarization and with a central
component surrounded by at least three outriders. The PA profile is
broken twice by orthogonal jumps, which coincide with the minima of the
linear polarization; this has also been noted in a separate analysis by \cite{kj06}. The RM profile is quite noisy, owing to the low
linear polarization, but flat, on average, across the entire pulse.

\subsection{Fit Accuracy Across the Band}
In order to check for the presence of any frequency-dependent, systematic effect that could cause a change of the PA across the frequency band that is different to the $\lambda^2$ law of Faraday rotation, we examined in more detail the fits that produced the calculated RMs for Vela, J1243$-$6423 and J1326$-$5859. 
Fig.~\ref{fig:0835rmfits}, \ref{fig:1243rmfits} and
\ref{fig:1326rmfits} show the RM fits for the minimum- and maximum-RM
bins (denoted by arrows) that reside near the central part of the
profile of the three aforementioned pulsars. It is important to note
that the associated errors on the fitted RMs, being only statistical,
would be underestimated if there was an additional, systematic,
non-Faraday component that affects the PAs. In order to reveal the
presence and magnitude of a possible systematic discrepancy between
the fitted function and the data, for each fit we also plotted the fit
residuals and their distribution (shown below the fit plots, in
Fig.~\ref{fig:0835rmfits}, \ref{fig:1243rmfits} and
\ref{fig:1326rmfits}). The quality of the data and the goodness of the
fit are evident in these figures. Upon closer inspection, however, the
residual PAs of all three pulsars, but mostly J1243$-$6423 and
J1326$-$5859, show evidence of some weak, sinusoidal fluctuations around the mean. It is difficult to attribute some significance to this, especially given
the Gaussian-like distributions of the residuals. The conclusion
therefore is that, if there is indeed a physical mechanism that affects the PA rotation, apart from the interstellar Faraday effect, we are unable to distinguish it from the dominant $1/f^2$ law.

\subsection{DM errors}
We also investigated the possibility that the RM variations observed are
caused by erroneous estimation of the DM, with, say, a $\Delta {\rm DM}$
offset from the true value. Such an offset is translated into a
dispersive delay between the pulse arrival time at a reference
frequency $f_0$ (usually the centre frequency of the band) and that at
frequency $f=f_0+\Delta f$, where $\mid\Delta f\mid \ll f_0$. In that
case, the delay in ms is given by

\begin{equation}
\label{eq:ddelay} 
\Delta t = 8.3\times 10^6 {\rm ms} \ \ \frac{\Delta f}{f_0^3}\Delta {\rm DM}
\end{equation} 
which also corresponds to a phase offset in the pulse window, $\Delta
\phi=\Delta t/P$, with $P$ the pulsar period.

If this phase offset is not accounted for, the calculation of the PA
in each frequency channel will correspond to a pulse longitude that is
shifted with respect to that at the reference frequency; this may
introduce a frequency-dependent shift in the PA of any given phase
bin. 
In order to eliminate this as the cause of the observed RM
variations, we de-dispersed our data using the published DM value, as well as two other values that were $5\sigma$ away, either side of the published one, and generated again the RM profile in each case. The published DMs for our pulsars are shown in Table~\ref{tab:pulprop}. 
By checking the resulting RM profiles for all pulsars, we concluded that there is no way of recovering a flat RM curve for any reasonable departure from the published DM. We therefore rule out DM errors as a possible cause of RM variation.

\section{Discussion}
Our analysis reveals that the RM profiles of pulsars with large RM
variations are diverse: they can be as simple as the `S'-shaped
profiles of Vela and PSR J1056$-$6258 or as complex as those of PSR
J1243$-$6423 and PSR J1644$-$4559. The peak-to-peak variation of those
pulsars' RMs across their profiles was found to be as little as a few
rad m$^{-2}$ (e.g. Vela in 2006) but also as large as 100 rad m$^{-2}$,
for PSR J1056$-$6258. Despite the hints of correlation between
individual features in the flux and PA profiles and those in the RM profiles (see e.g.~PSR J1326$-$5859), in most cases these correlations
seem coincidental. However, a general observation is that the greatest RM fluctuations across the pulse profile seem coincident with the steepest gradients of the PA profile, whereas pulsars with flat PA profiles show little or none at all RM variation. Good examples that support this argument are PSR J1243$-$6423 and PSR J2048$-$1616, whose RM departure from the mean is maximised around the inflexion point of the PA swing; in contrast, PSR J0907$-$5157 and PSR J1057$-$5226, having flat PA profiles, show insignificant RM variations.   

The subset of pulsars with low RM variations adheres to the
theoretical predictions. More than half of those pulsars have
phase-resolved RMs that are scattered around their respective
published RM value, thus showing convincingly longitude-independent RM
values across the pulse. However, there are cases like those of PSR
J1157$-$6224, PSR J1253$-$5820, PSR J1359$-$6038 and PSR J1705$-$1906,
where the RM scatter across the pulse is clearly not
Gaussian: these profiles show a clear structure but with a small RMS
spread. An interesting case is that of PSR J0742$-$2822, where
although no significant RM variations exist across the profile, the
longitude-averaged RM across the pulse is higher in 2004 data than in
2006. Nevertheless, one has to be careful when drawing the above
conclusions; the main statement is that any RM variations with pulse
phase are smaller than the statistical error of the measurement.

We have demonstrated that the RM variations observed are not due to measurement errors, and that the PA rotation in all cases follows the
$\lambda^2$ law of Faraday Rotation. In the following we consider three possible explanations: (a) Faraday Rotation within the pulsar magnetosphere, (b) the superposition and frequency dependence of quasi-orthogonal
polarization modes, and (c) interstellar scattering and its effects on the polarization.

\subsection{Magnetoshperic Faraday Rotation}

Faraday Rotation through the ionised ISM has the effect of rotating
the PA with frequency. Our data, however, comprise of full sets of the
four Stokes parameters, $I$, $Q$, $U$ and $V$, so we can observe the frequency
dependence of the entire polarization vector
($\hat{\boldsymbol{P}}=(Q\hat{\boldsymbol{i}}+U\hat{\boldsymbol{j}}+V\hat{\boldsymbol{k}})/\sqrt{Q^2+U^2+V^2}$)
as a function of frequency, for specific high-s/n phase bins. An
example for one phase bin of PSR
J1326$-$5859 is shown in Fig.~\ref{fig:poincare}, where the tip of the Poincar\'e vector is plotted across
32 frequency channels. This plot demonstrates that, together with the
pronounced effect of PA rotation (around the vertical axis), a small
and rather systematic change in the fraction of circular polarization
can be seen. We obtained similar plots for four additional pulsars in
our sample that displayed large RM variations: namely Vela, PSR
J1243$-$6423, PSR J1644$-$4559 and PSR J2048$-$1616. In nearly all
cases, we found that the trace of $\hat{\boldsymbol{P}}$ on the
Poincar\'e sphere does not strictly follow paths of constant latitude,
as in Fig.~\ref{fig:poincare}.

The wave modes in the cold and ionised ISM are purely circularly
polarised. Therefore, the phase offset introduced by propagation
through this medium results in the well-documented changes of the
PA. On the contrary, in a medium of highly relativistic charged
particles, the natural modes are close to linearly polarised (\nocite{saz69}Sazonov 1969; \nocite{mel97}Melrose 1997a). As a
consequence, different phase offsets between the modes will result in
a conversion between linear and circular polarization. In a pulsar's
magnetosphere, the observed degrees of circular polarization indicate
that the natural wave modes are elliptically polarised, which suggests
that propagation through this medium will generate some frequency-dependent conversion between linear and circular polarization. An
extensive discussion on this can be found in \cite{km98}. 

If such a generalised Faraday Rotation is taking place within the
pulsar magnetosphere and giving rise to RM variations as well as
changing the degree of circular polarization in the observed
frequency band, we expect the greatest change in $V$ to coincide with
the greatest variations in RM.  
In order to test this, we first calculated the fraction of circularly
polarised intensity, $V/I$, for each frequency channel across the
band. The division by $I$ removes any fluctuations of $V$ caused by
scintillation through the interstellar medium, whereas any change in $V$
that originates in the magnetoshpere should remain
unaffected. Subsequently, we fitted a linear function, $V/I \propto
f$, to the data and calculated the change in $V/I$ across the band,
based on that function. This was done to get a good estimate of the
total effect while minimizing the influence of instrumental noise. We
found linear fits to describe the data sufficiently well.
The resulting profiles of $\Delta(V/I)$ as a function of longitude,
for Vela, PSR J1243$-$6423, PSR J1326$-$5859, PSR J1644$-$4559 and PSR
J2048$-$1616 are shown in Fig.~\ref{fig:VDVprofiles}.  A side-by-side
comparison of the RM and $\Delta(V/I)$ profiles, shown in
Fig.~\ref{fig:VDVprofiles}, makes clear that none of the pairs of profiles seems correlated. 

There is certainly little evidence that the maximum RM deviation from the mean and the maximum change in circular polarization across the band occur at the same pulse phase. Therefore, to first order, the assumption that the observed RM variations are caused by magnetospheric Faraday rotation is deemed unlikely.

\subsection{Quasi-orthogonal Emission Modes}
Another factor that could potentially affect the measured RM is the
{\em incoherent}\/ superposition of quasi-orthogonal modes of
polarization (see e.g.~\nocite{rbr+04}Ramachandran et al. 2004). The superposition of strictly
orthogonal modes results in an observed PA of the strongest of the two
modes, regardless of the ratio of the intensity of the modes
(e.g.~\nocite{brc76}Backer et al 1976; \nocite{crb78}Cordes et al. 1978). The reason is that orthogonal modes are represented by
anti-parallel vectors on the Poincare sphere; regardless of their
relative lengths, the addition is always pointing in the direction of
either of the two modes. This is not the case for non-orthogonal
modes: in this case, the sum of the mode vectors points at an angle
that depends on the degree of non-orthogonality and the relative
intensity of the modes. As a consequence, if the ratio of the mode
intensities changes as a function of frequency, the observed PA will
also rotate. The question is under what conditions this could
resemble Faraday Rotation.

In the pulsars where significant RM variations are seen, the
variations often exceed 20 rad m$^{-2}$ across the pulse profile. At
the frequency and bandwidth of these observations, this corresponds to
a variation in PA by $\sim 18.5^\circ$. Therefore, in order for
quasi-orthogonal polarization modes to present a possible explanation
for the observed RM variations, they should be capable of inducing
such a PA rotation.

From polarization studies of 18 pulsars in the European Pulsar Network
(EPN) archive, \cite{sse+06} conclude that the ratio of the average
intensity of the orthogonal modes for those pulsars follows a power
law with spectral index that does not typically exceed
$\gamma \sim -0.5$. Assuming linear modes ($L_1$ and $L_2$; the
degree of circular polarization is generally very low), then at a
reference frequency $f_0$ the mode intensity ratio is
$R_0=L_1/L_2$, and the form of this power law is
\begin{equation} 
\label{eq:opmpl} 
R=R_0\left(\frac{f}{f_0}\right)^{-0.5} 
\end{equation} 

For given values of $R_0$, we can compute $R$ across the frequency band
we are interested in: in this case, between 1300 and 1556 MHz. We have
numerically computed the total rotation of the PA across the band (hereafter
$\Delta$PA), for different values of $R_0$ and different angles of
non-orthogonality, $\phi$. The latter angle is defined on the Poincar\`e
sphere as the supplementary to the angle between the two modes. Figure
\ref{fig:dpafunc} shows $\Delta$PA versus $\phi$, for various values of $R_0$,
which equate to certain degrees of linear polarization. The plot on
the left shows examples where $R_0\geq 1$ as opposed to $R_0\leq 1$
for the plot on the right. Both plots show a set of lines
corresponding to the same degrees of polarization , $L$\%, which is
approximately $\mid L_1-L_2\mid/I$.  Two things are immediately
evident: first, the greatest rotation of PA happens for small values
of $\phi$; second, and most important, the only possibility that
$\Delta$PA$>15^\circ$ occurs when the fractional polarization remains under
10\% and only in the specific case where $R_0>1$, i.e. the case where
the mode with the steepest spectral index is brightest at the lowest
frequency $f_0$.  In our data, there are clear examples of high $\Delta$RM
and high $L$\%, the most prominent being Vela and J1326$-$5759.  It is
therefore unlikely that non-orthogonal polarization modes provide a
general explanation for the RM variations in the data.

\subsection{Scattering}
As was mentioned earlier in this paper, some of the pulsars we studied
show signs of interstellar scattering in their flux profiles: vivid
examples are PSR J1056$-$6258, justified by its high DM of 320.3 pc
cm$^{-3}$, and PSR J1644$-$4559, with a DM of 478.8 pc
cm$^{-3}$ --- the highest in
our sample. We checked for a possible correlation between the amount
of interstellar scattering and the magnitude of the observed RM
variations. The effect of interstellar scattering is usually
quantified by the scattering timescale, $\tau_{\rm sc}$. Assuming 
a Kolmogorov spectrum for the density fluctuations along the
sightline to the pulsar, $\tau_{\rm sc}$ has a roughly quadratic 
relation to DM for low-DM pulsars (i.e.~$\tau_{\rm sc}\propto {\rm DM}^{2}$) 
but can be much steeper for high-DM pulsars: i.e.~$\tau_{\rm sc}\propto {\rm DM}^{4}$ (e.g.~\nocite{sut71}Sutton 1971; \nocite{ric77}Ricket 1977). More recently, based on a global fit to 98 scattering times of high-DM pulsars, \cite{bcc+04} showed that a parabolic curve of the form

\begin{equation} 
\label{eq:tausc}
\log\tau_{\rm sc}=-6.46+0.154\log{\rm DM}+1.07(\log{\rm DM})^2-3.86\log{f_0}
\end{equation} 
where $f_0$ is the observing frequency, is a reasonable description of the trend of $\tau_{\rm sc}$ as a function of DM. Hence, as long as we are refering to the same observing frequency, and since there are only a few available measurements of $\tau_{\rm sc}$ for our pulsar sample (see Table~\ref{tab:pulprop}), the DM is a convenient alternative to the amount of scattering a pulsar's emission is subject to. 

We have quantified the magnitude of the RM variations for those
pulsars whose RM profiles show signs of a longitude-dependent
structure, in terms of the peak-to-peak distance of their RM profile,
$\Delta {\rm RM}_{\rm pp}$. Fig.~\ref{fig:dRMmaxDM} shows a scatter
plot of $\Delta {\rm RM}_{\rm pp}$ against DM for 9 pulsars. Although
there is perhaps no clear correlation, it is striking that the pulsars
that typify high RM variations are also high-DM pulsars. This includes
the most prominent examples of RM variation with pulse phase, such as
PSR J1243$-$6423, PSR J1326$-$5859 and PSR J1644$-$4559. In a recent
paper, \cite{kar09} showed that small amounts of scattering and
orthogonal polarization modes of emission in the pulsar magnetosphere
can result in a frequency dependent PA that can be falsely attributed
to Faraday Rotation. This explanation is extremely simple and may well
be the key to understanding the high-DM examples. It is also
strengthened by the observation that most of the low-DM pulsars show
relatively smooth PA profiles, devoid of the bumps and wiggles seen in
the higher DM profiles. The overall correlation in
Fig.~\ref{fig:dRMmaxDM} is perhaps reduced by the fact that RM
measurements can only be made for pulse phases with significant linear
polarization.

\section{Conclusions}

We have shown phase-resolved RM profiles for 19 pulsars, 9 of which show
substantial RM variations across the pulse profile. These variations appear in
most cases to show coherent rather than noisy behaviour. The high s/n of our
data and our robust fitting technique have shown that whatever the cause of the
PA rotation with frequency, it is inseparable from the $1/f^2$ law of
interstellar Faraday Rotation within our observed band. 

We have considered a
number of possibilities that may result in the observed phenomena. For example,
we have ruled out the possibility that de-dispersion with a slightly erroneous
DM may be responsible for the data here. We have discussed the possibility that
the RM variations are of magnetospheric origin. Two cases were examined, that of
generalised Faraday Rotation due to the relativistic plasma in the pulsar
magnetospheres, and the existence and superposition of quasi-orthogonal modes of
polarization. 

In the case of generalised Faraday Rotation, we examined whether
pulse phases with the highest rotation of the PA also showed the highest change
of the fractional circular polarization. Our first-order investigation turned
out negative results, although it is true that the theory for generalised
Faraday Rotation is not complete and does not provide us with good handles on
the rate at which polarization rotations occur with frequency. 

As concerns
quasi-orthogonal modes, we find that it is impossible to obtain a large rotation
of the PA within our observing frequency band and observe large degrees of
linear polarization at the same time. Our analysis indicates that there are
specific cases of the data that can be explained by this process, but it
certainly does not form a general framework for explaining RM variations across
the pulse. 

Finally, this work supports the ideas presented in \cite{kar09}, that interstellar scattering can distort the polarization in a frequency-dependent way. This may resemble Faraday Rotation and will generally be
pulse-phase dependent. It is important to emphasise that the possibility of
magnetospheric Faraday Rotation is not totally eliminated by this work; we
merely conclude that it is an unlikely explanation for RM variations of the
observed magnitude.

There is a possibility that both the magnetospheric mechanisms and
interstellar scattering are responsible for the observed RM
variations, along with other possibe mechanisms which we have not
considered. Whichever the explanation, certain care must be taken
during RM-determination experiments from pulsars to correctly
identify the amount of PA rotation that is due to the iono-magnetised
interstellar medium. Especially for the studies of the planar component 
of the Galactic magnetic field, where pulsars with high DMs and RMs are used,  
the identification of the ISM-induced Faraday rotation and the correct 
calculation of the associated RM plays an important role in our efforts to model the interstellar magnetic fields. Hence, in light of the RM profiles presented herein, pulsar-RM determination from specific, high-s/n pulse bins can lead to considerable errors.

\vspace{0.1cm}

\section*{Acknowledgements} 
\noindent The Australia Telescope is funded by the Commonwealth of
Australia for operation as a National Facility managed by the CSIRO.
We would like to thank Andrew Lyne, Don Melrose and Roy Smits for useful discussions and comments on this manuscript.

\vfill

\pagebreak

\clearpage

\renewcommand{\theequation}{A-\arabic{equation}}
  % redefine the command that creates the equation no.
\setcounter{equation}{0}  % reset counter 
%\section*{APPENDIX}  % use *-form to suppress numbering

\clearpage

\begin{table}
\caption{\label{tab:pulprop} Tabulated properties of the 19 pulsars we analysed in this work. Columns 7 and 8 show the pulse-averaged fraction of the linearly and circularly polarised intensity, respectively. The last column shows the magnitude of the interstellar pulse broadening as a percentage of the pulse's FWHM; this is only known for 6 pulsars in our sample.}  
\centering 
\footnotesize
\begin{tabular}{@{}lrrrrrrrrrr} 
\hline 
         PSR      &     $P$            &       DM           &  $\sigma_{\rm DM}$  &   RM$_{\rm pub}$  &  $\sigma_{\rm RM}$  &  $L/I$  & $V/I$  &  $\Delta{\rm RM}_{\rm pp}$  &  $S_{1400}$  &  $\tau_{\rm sc}/{\rm W}_{50}$     \\ 
                  &    [s]             &  [pc cm$^{-3}$]    &   [pc cm$^{-3}$]    &   [rad m$^{-2}$]  &  [rad m$^{-2}$]     &   [\%]  &  [\%]  &  [rad m$^{-2}$]             &  [mJy]       &  [\%]                                \\ 
\hline 

   J0134$-$2937   &   0.1369           &          21.806    &       0.006         &     15        &          3              &  41   &  19      &   0                         &   2.4        &      --                                \\
   J0738$-$4042   &   0.3749           &          160.8     &       0.7           &     12.1      &          0.6            &  29   &   6      &   0                         &  80          &      --                                \\
   J0742$-$2822   &   0.1667           &          73.782    &       0.002         &    148.5      &          0.6            &  88   &   3      &   0                         &  15          &     0.14                               \\
   J0835$-$4510   &   0.0893           &          67.99     &       0.01          &     30.4      &          0.6            &  88   &   7      &   13                        &  1100        &     3                                   \\
   J0907$-$5157   &   0.2535           &         103.72     &       --           &  $-$23.3      &          1              &  32   &   4      &   0                         &  9.3        &      --                                     \\
   J0942$-$5552   &   0.6643           &         180.2      &       0.5           &  $-$61.9      &          0.2            &  33   &   8      &   15                        &  10        &       --                                    \\
   J1056$-$6258   &   0.4224           &         320.3      &       0.6           &      4        &          2              &  50   &   2      &   100                       &  21        &       --                                    \\
   J1057$-$5226   &   0.1971           &          30.1      &       0.5           &     44        &          2              &  86   &   4      &   0                         &  --        &       --                                    \\
   J1157$-$6224   &   0.4005           &         325.2      &       0.5           &    508.2      &          0.5            &  26   &  12      &   0                         &  5.9        &      --                                    \\
   J1243$-$6423   &   0.3884           &         297.25     &       0.08          &    157.8      &          0.4            &  20   &  13      &   52                        &  13        &       --                                    \\
   J1253$-$5820   &   0.2554           &         100.584    &       0.004         &     31        &          5              &  57   &   6      &   0                         &  3.5        &      --                                    \\
   J1326$-$5859   &   0.4779           &         287.30     &       0.15          & $-$579.6      &          0.9            &  32   &  17      &   37                        &  9.9        &     38                                     \\
   J1359$-$6038   &   0.1275           &         293.71     &       0.14          &     33        &          5              &  73   &  15      &   13                        &  7.6        &      --                                    \\
   J1453$-$6413   &   0.1794           &          71.07     &       0.02          &  $-$18.6      &          0.2            &  24   &   6      &   31                        &  14        &       --                                   \\
   J1644$-$4559   &   0.4550           &         478.8      &       0.8           & $-$617        &          1              &  17   &   3      &   25                        &  310        &     136                                     \\
   J1705$-$1906   &   0.2989           &          22.907    &       0.003         &   $-$9        &          4              &  72   &  27      &   0                         &  8        &        --                                     \\
   J1745$-$3040   &   0.3674           &          88.373    &       0.004         &  $-$19.2      &          1              &  49   &   5      &   11                        &  13        &       --                                     \\
   J1807$-$0847   &   0.1637           &         112.3802   &       0.0011        &   166         &          9              &  23   &   9      &    0                        &  15        &      0.03                                      \\
   J2048$-$1616   &   1.9615           &          11.456    &       0.005         &  $-$10.2      &          0.8            &  37   &   4      &   17                        &  13        &      $10^{-6}$                                      \\

\hline 
\end{tabular} 
\end{table}

\setcounter{figure}{0}

\begin{figure}\begin{center} 
\vspace*{10pt}
\includegraphics[width=0.8\textwidth]{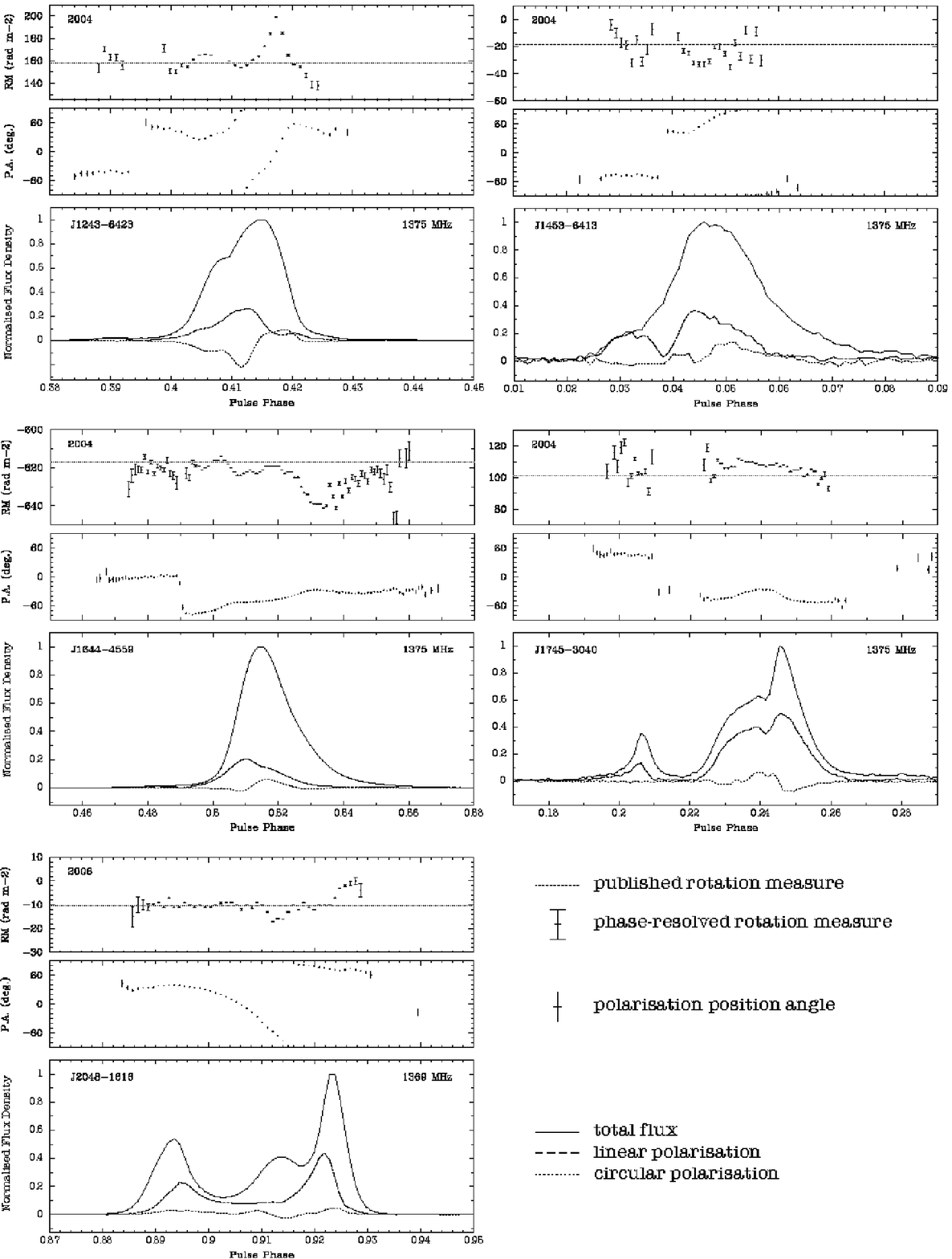}
\caption{\label{fig:largeRMvar2} Continued}
\end{center}\end{figure}

\begin{figure}\begin{center} 
\vspace*{10pt}
\includegraphics[width=0.85\textwidth]{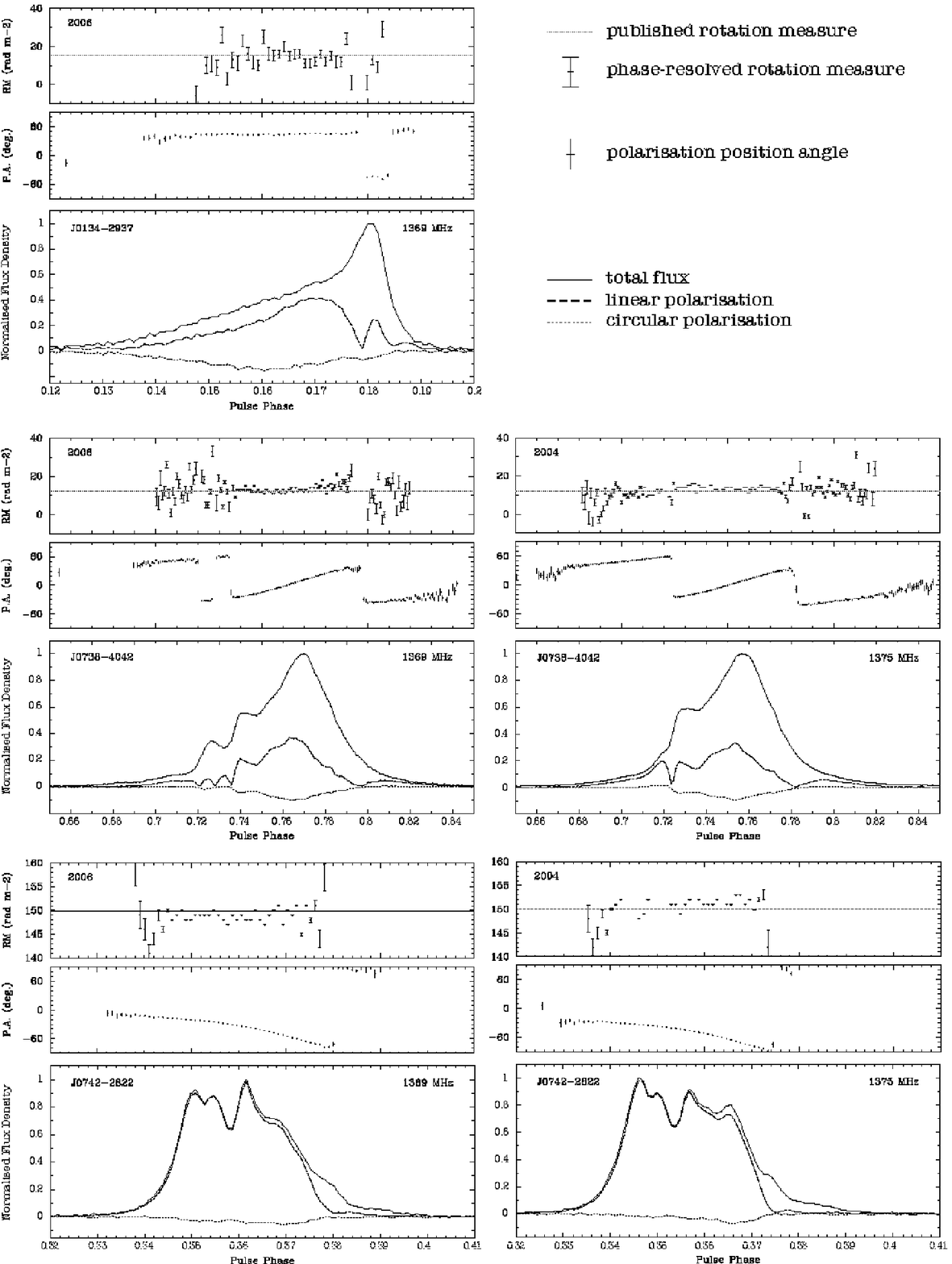}
\caption{\label{fig:smallRMvar1} Profiles of the RM as a function of pulse longitude (top panels), for 10 pulsars, observed at 20 cm with the Parkes telescope, that exhibit small or no RM variations across their pulse profile. The respective year of observation is shown at the top-left corner of each pulsar's RM profile. For two pulsars, namely PSR J0738$-$4042 and PSR J0742$-$2822, we had data from two observing sessions, in 2004 and 2006. See Fig.~\ref{fig:largeRMvar1} for more details.}
\end{center}\end{figure}

\setcounter{figure}{1}

\begin{figure}\begin{center} 
\vspace*{10pt}
\includegraphics[width=0.75\textwidth]{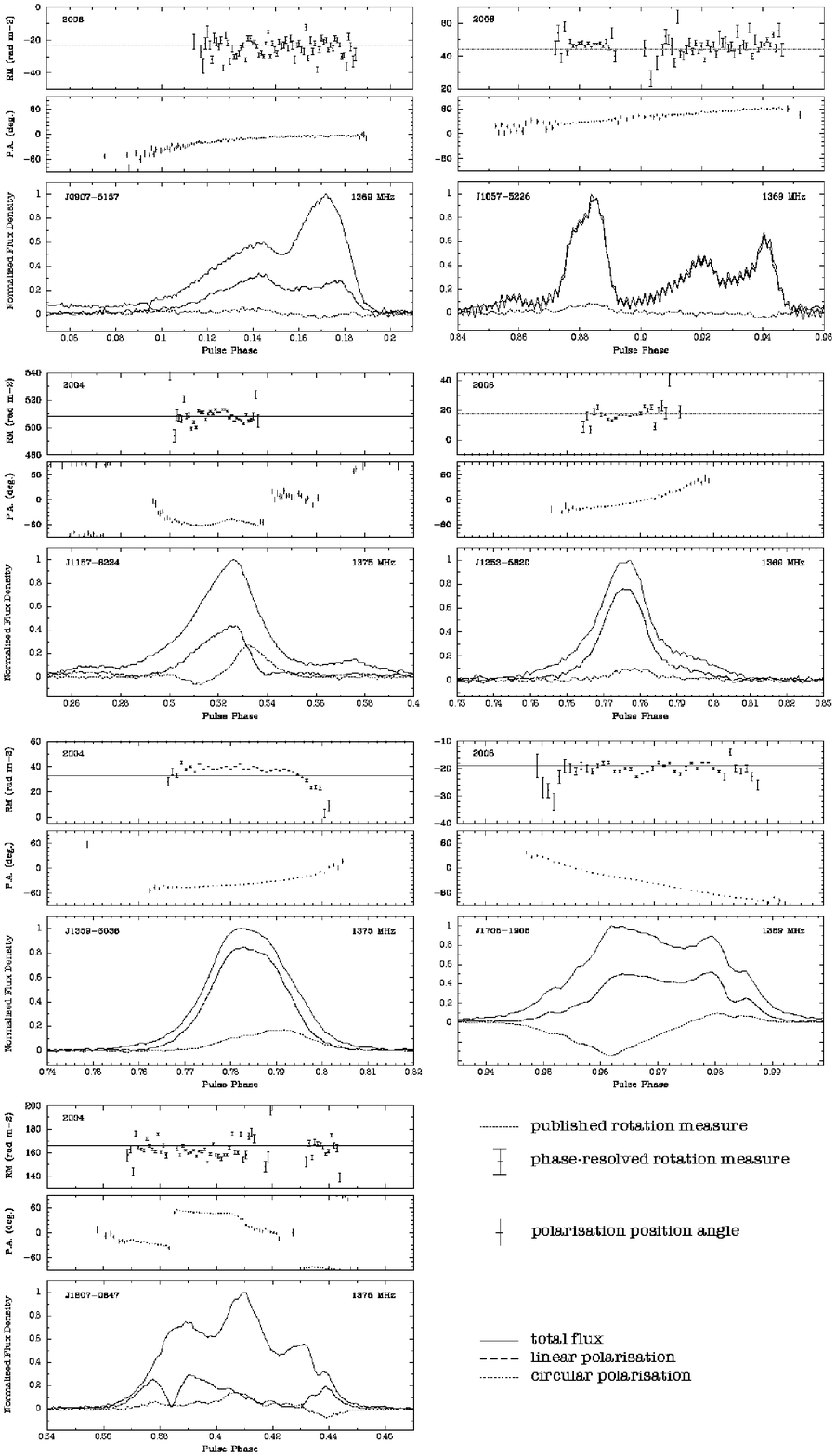}
\caption{\label{fig:smallRMvar2} Continued}
\end{center}\end{figure}

\begin{figure}\begin{center}  
\vspace*{10pt} 
\includegraphics[width=0.98\textwidth]{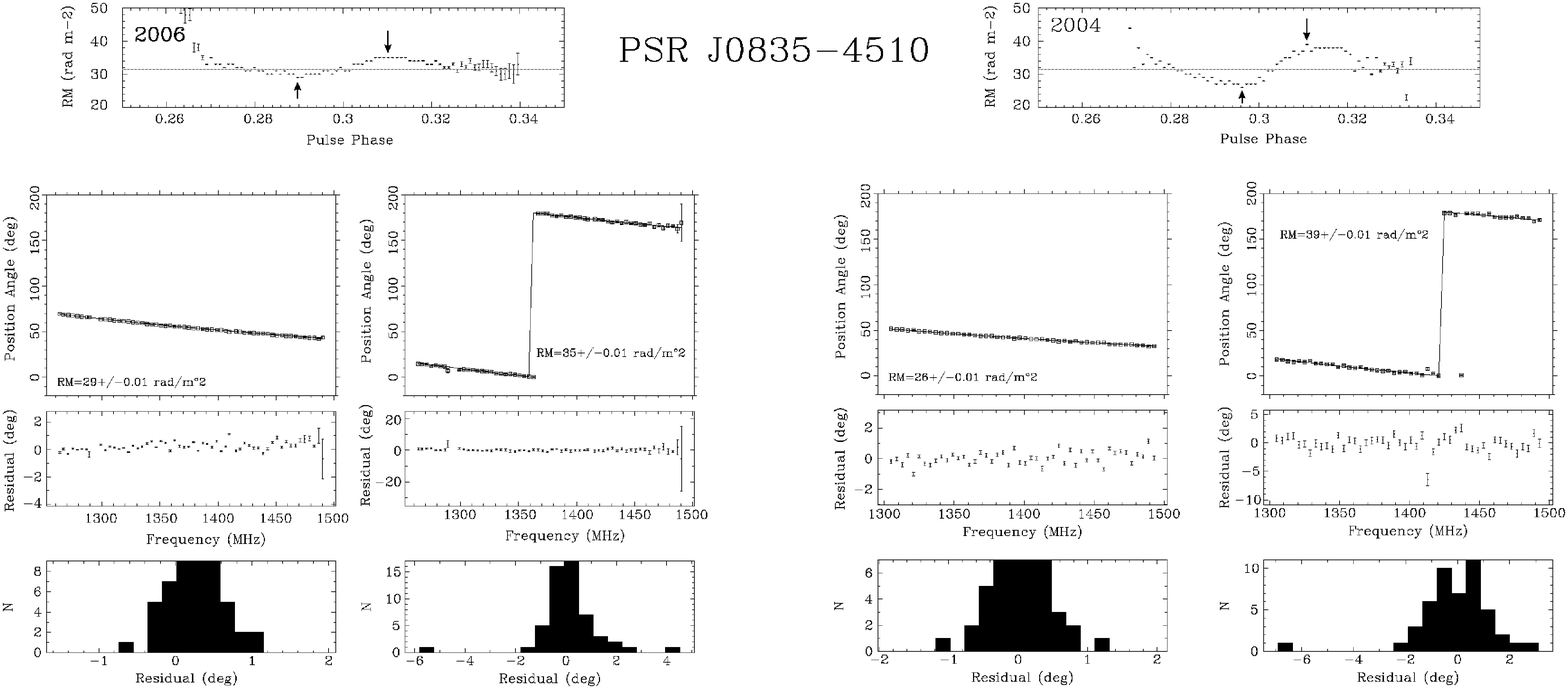} \caption{\label{fig:0835rmfits} RM fits to the PAs corresponding to
the maximum- and minimum-RM values in the central part of Vela pulsar's RM profile. The set of plots to the left of the pulsar's J2000 name (J0835$-$4510)
correspond to the 2006 data; the ones to right of the legend correspond to data collected in 2004. The top plots in each set show the RM variation across the
pulse; below them, we show the RM fits corresponding to the minimum (left) and maximum RM (right) in the RM profile. The minimum- and maximum-RM values selected for the fits are indicated with black arrows in the respective RM profiles. Below each fit we show the fit residuals in degrees, and at the very bottom the distribution of those residuals is shown.} 
\end{center}\end{figure}

\begin{figure}\begin{center} 
\vspace*{10pt}
\includegraphics[width=0.47\textwidth]{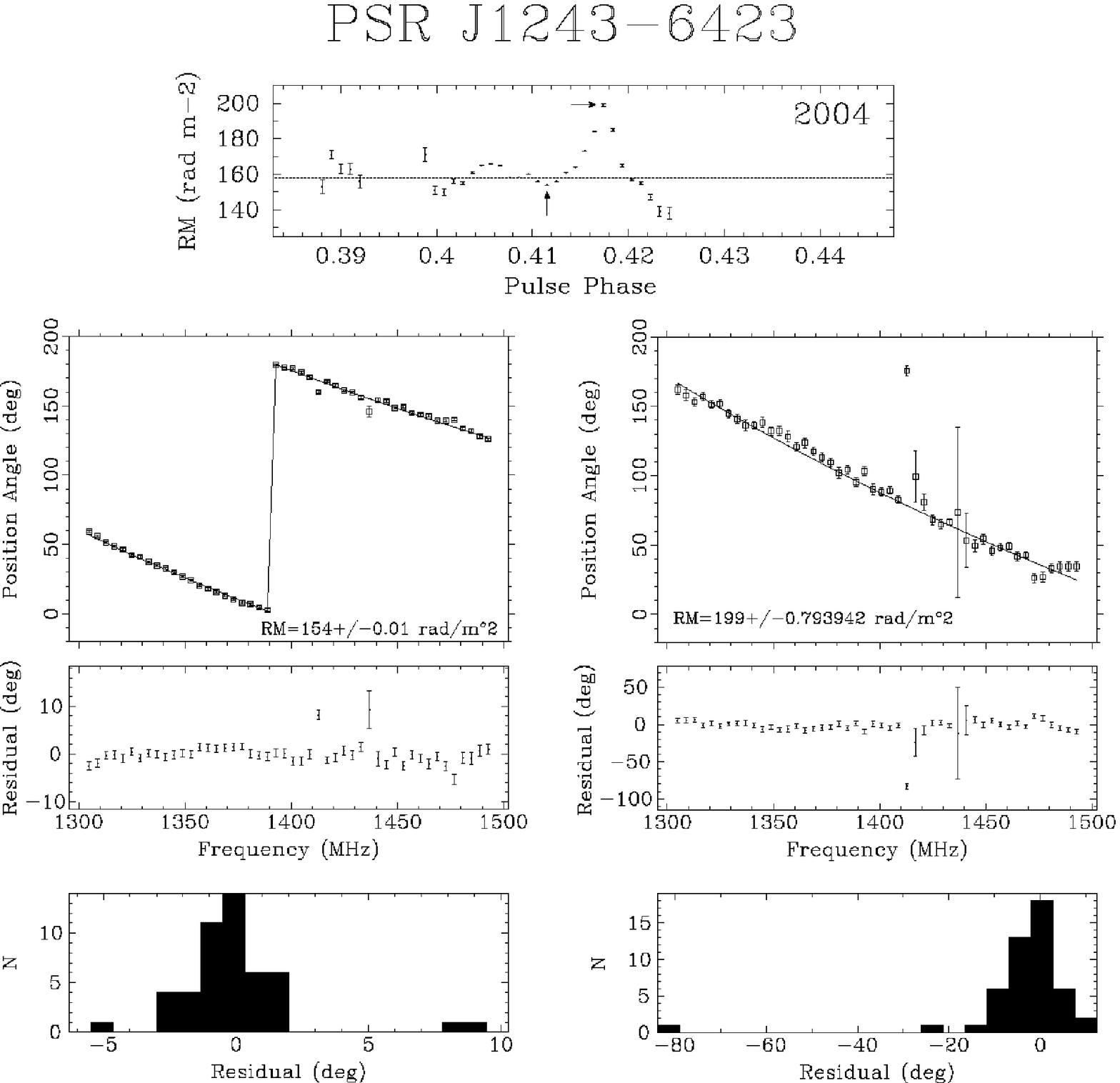}
\caption{\label{fig:1243rmfits} RM fits for J1243$-$6423. See Fig.~\ref{fig:0835rmfits} for details.}
\end{center}\end{figure}

\begin{figure}\begin{center} 
\vspace*{10pt}
\includegraphics[width=0.98\textwidth]{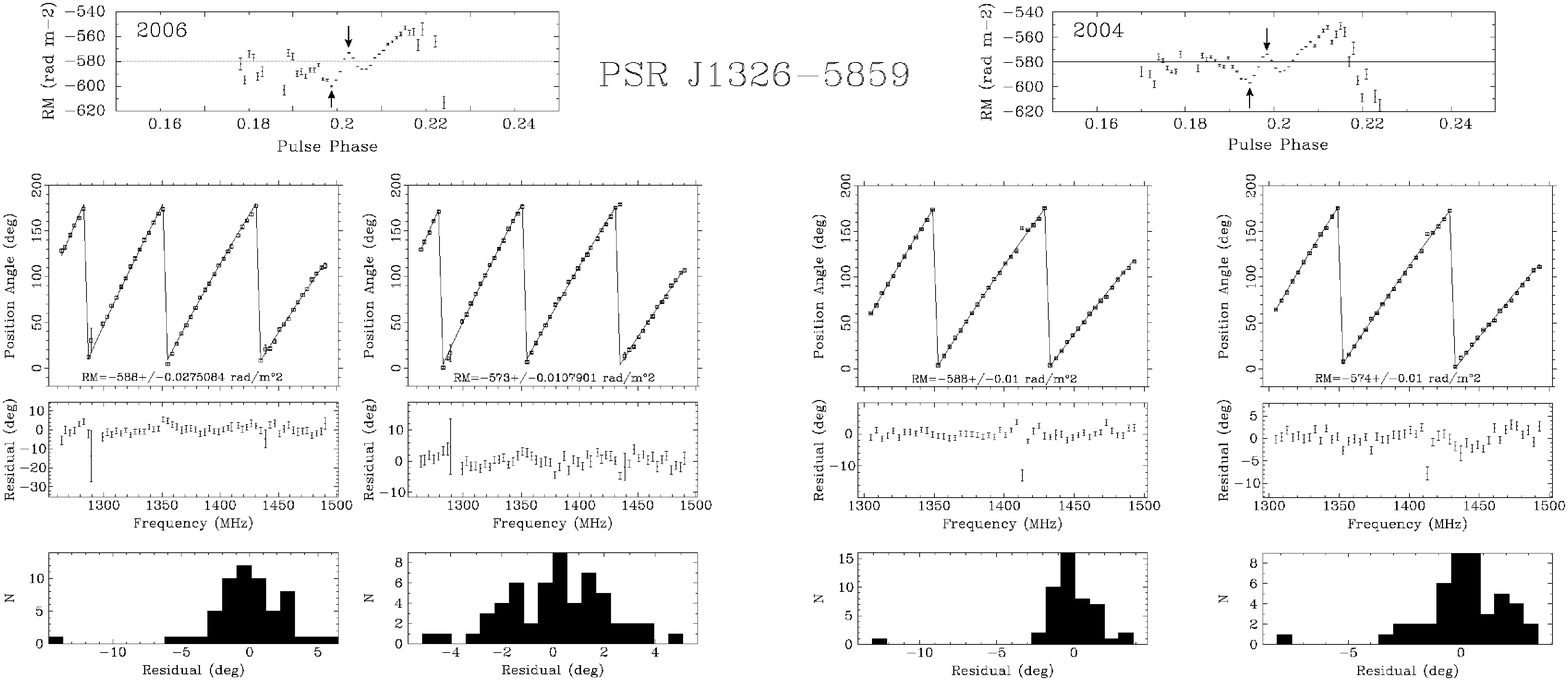}
\caption{\label{fig:1326rmfits} RM fits for J1326$-$5859. See Fig.~\ref{fig:0835rmfits} for details.}
\end{center}\end{figure}

\begin{figure}\begin{center} 
\vspace*{10pt}
\includegraphics[width=0.55\textwidth]{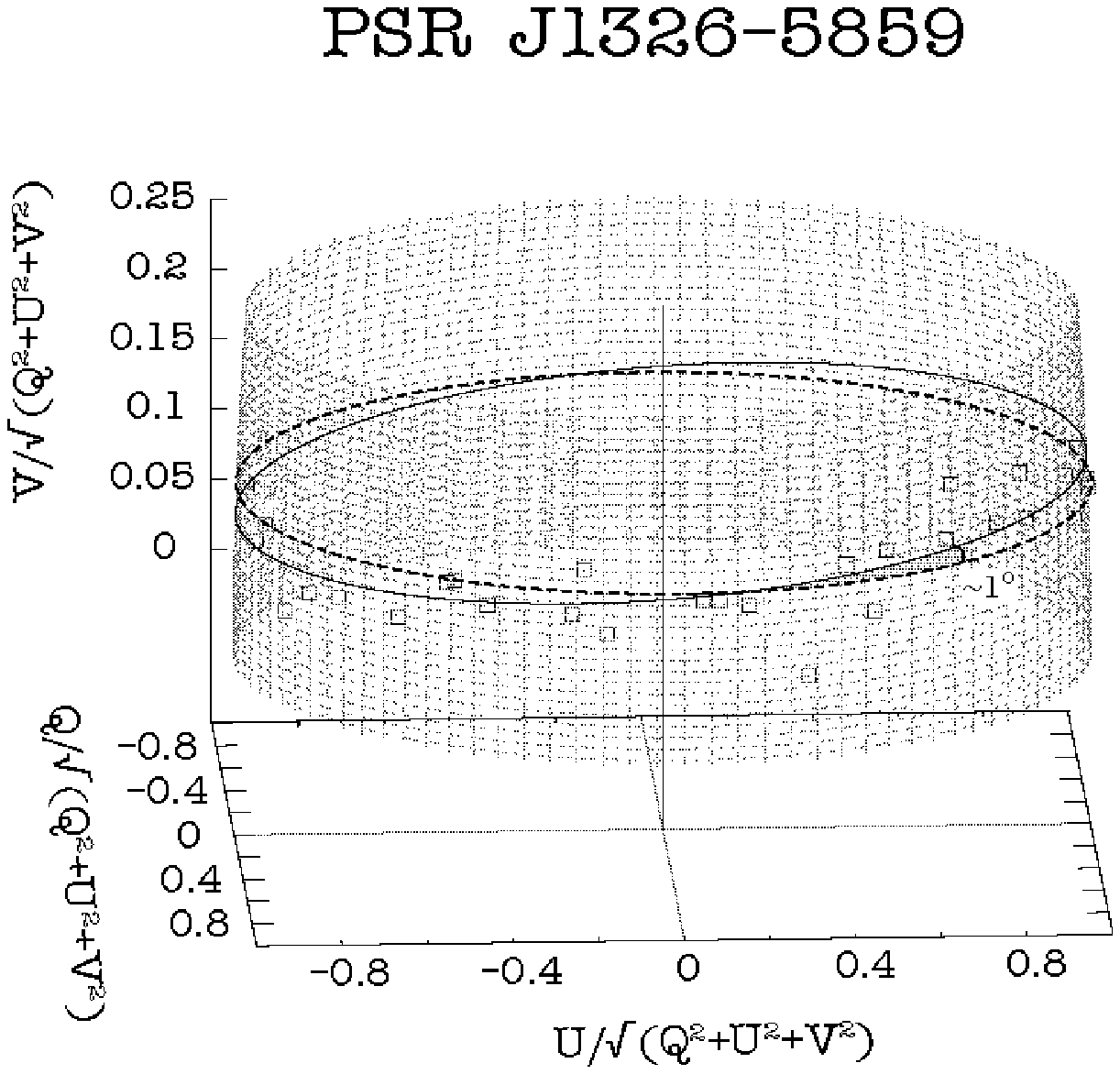}
\caption{\label{fig:poincare} The rotation of the Poincar\'e vector, $\hat{\boldsymbol{P}}$, as a function of frequency, across the frequency band, for PSR J1326$-$5859. The 3D representation of the rotation is zoomed into the latitude range which contains the data. The data points were best fitted with a circle of inclination $1.3^\circ\pm 0.6^\circ$ (solid black line), whereas a flat circle of zero inclination (dashed black line) gave a worse fit.}

\end{center}\end{figure}

\begin{figure}\begin{center} 
\vspace*{10pt}
\includegraphics[width=1\textwidth]{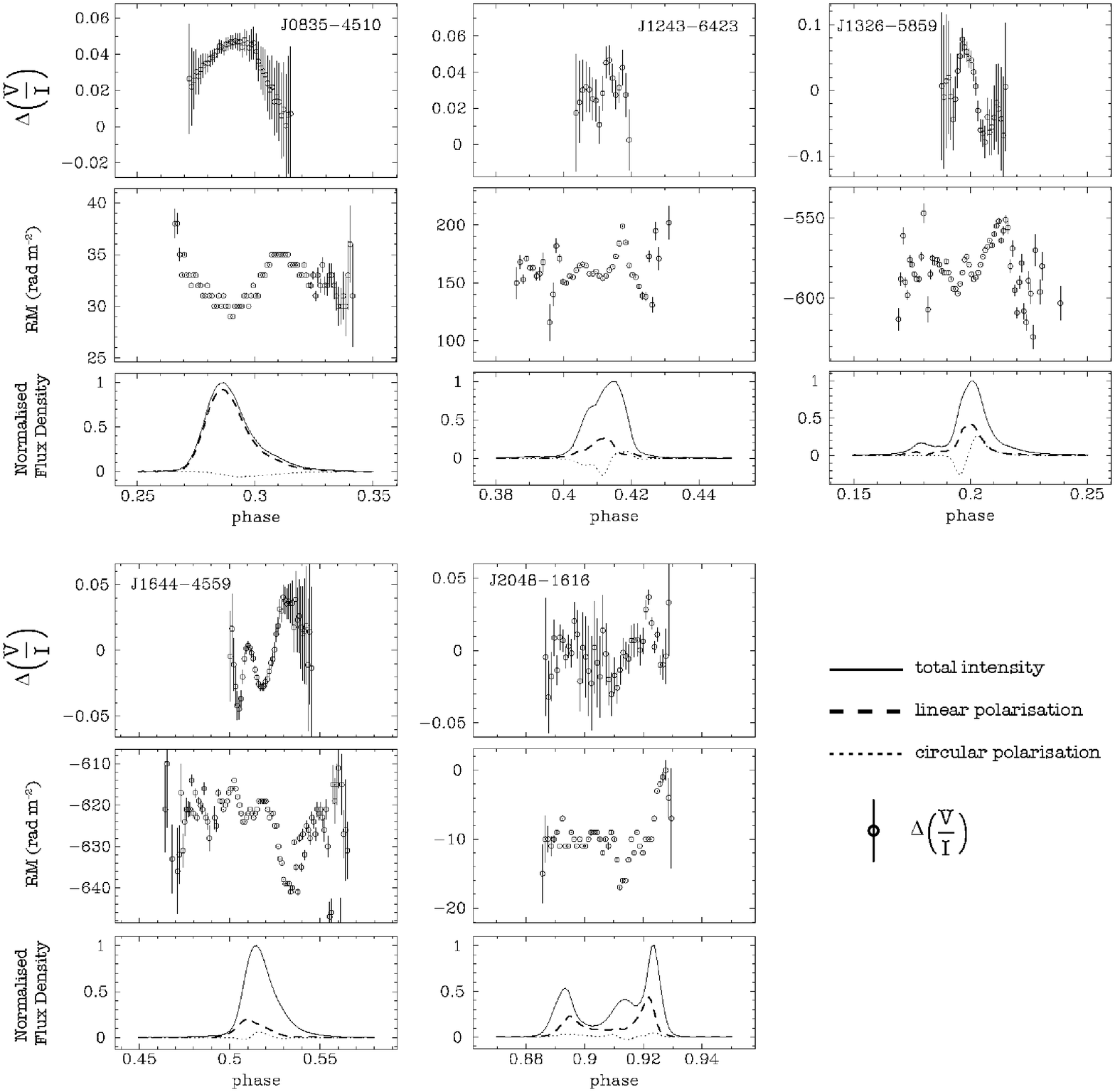}
\caption{\label{fig:VDVprofiles} Top panels: the change in the fraction of circularly polarised intensity across the band as a function of pulse longitude, for Vela, PSR J1243$-$6423, PSR J1326$-$5859, PSR J1644$-$4559 and PSR J2048$-$1616. For comparison, we have included, in the middle and bottom panels, the RM and polarization profiles of each pulsar.}
\end{center}\end{figure}

\begin{figure}\begin{center} 
\vspace*{10pt}
\includegraphics[width=0.47\textwidth]{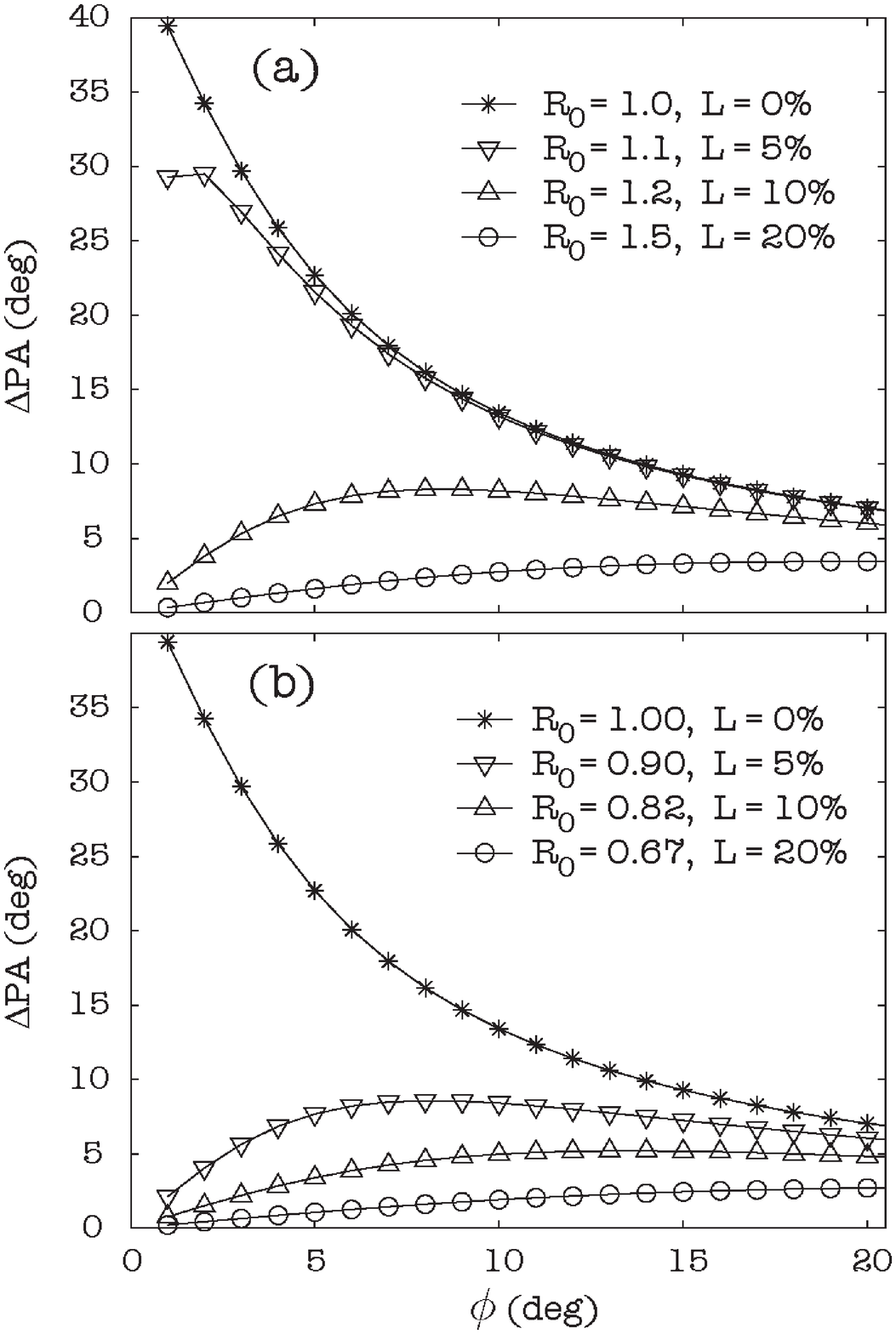}
\caption{\label{fig:dpafunc} The total PA difference between 1300 and
  1556 MHz due to the frequency evolution of the relative strength of
  quasi-orthogonal modes, as a function of the angle of departure from
  orthogonality, $\phi$, and for (a) increasing and (b) decreasing values of the initial
  relative intensity of the modes $R_0$, as indicated.}
\end{center}\end{figure}

\begin{figure}\begin{center} 
\vspace*{10pt}
\includegraphics[width=0.47\textwidth]{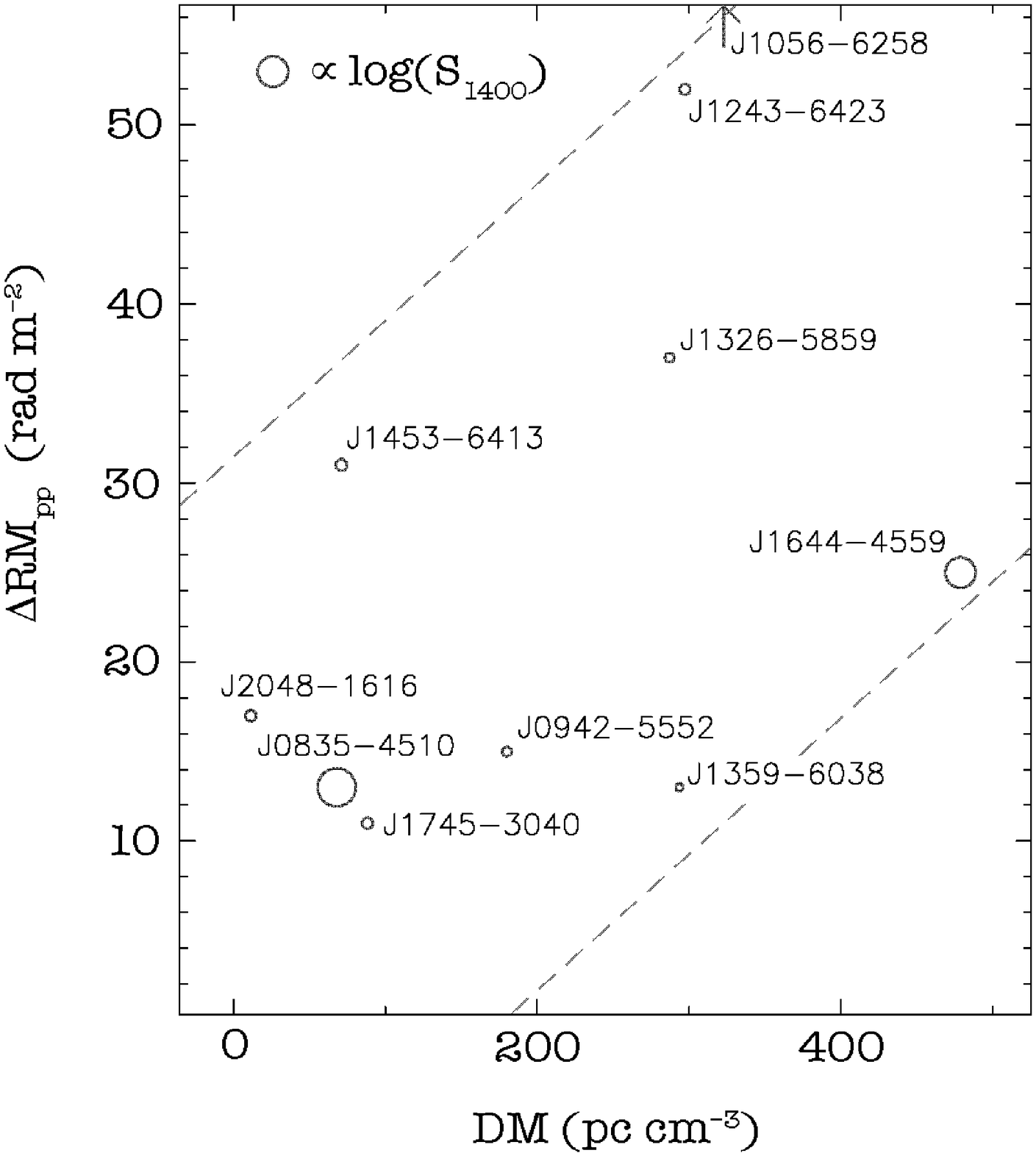}
\caption{\label{fig:dRMmaxDM} Scatter plot of the maximum, peak-to-peak RM difference ($\Delta {\rm RM}_{\rm pp}$) as a function of the pulsar DM,
measured across the profiles of the 9 pulsars that show large RM variations and PSR J1359$-6038$, for which a lesser degree of RM variation is visible. The arrow at the top of the frame indicates that the $\Delta {\rm RM}_{\rm pp}$ value measured for PSR J1056$-$6258 ($\sim 100$ rad m$^{-2}$) is off the plot's scale. The symbol size in this plot is proportional to the logarithm of the average flux at 1.4 GHz, $S_{1400}$. }
\end{center}\end{figure}

\end{document}